\documentclass[final,leqno,onefignum,onetabnum]{siamltex1213}

\usepackage{amsmath,amssymb,cite}
\usepackage{pifont}

\title{Convergence of direct recursive algorithm for identification of Preisach hysteresis model with stochastic input}

\author{D. Rachinskii\thanks{Department of Mathematical Sciences, The University of Texas at Dallas, USA
(\email{dmitry.rachinskiy@utdallas.edu}). The author acknowledges the support of NSF through grant DMS-1413223. Questions, comments, or corrections
to this document may be directed to that email address.} \and
M. Ruderman\thanks{Department of Computer Science and Engineering, Nagoya Institute of Technology, Japan (\email{ruderman.michael@nitech.ac.jp}).}}

\begin{document}
\maketitle
\slugger{siap}{xxxx}{xx}{x}{x--x}

\begin{abstract}
We consider a recursive iterative algorithm for identification of parameters
of the Preisach model, one of the most commonly used models of hysteretic input-output
relationships.
The classical identification algorithm due to Mayergoyz
defines explicitly a series of test inputs that allow one to find parameters
of the Preisach model with any desired precision provided that
(a) such input time series can be implemented and applied; and, (b) the corresponding output data
can be accurately measured and recorded. Recursive iterative
identification schemes suitable for
a number of engineering applications have been recently proposed
as an alternative to the classical algorithm.
These recursive schemes do not use any input design but rather rely on
an input-output data stream resulting from random fluctuations of the input variable.
Furthermore, only recent values of the input-output data streams are available
for the scheme at any time instant. In this work, we prove exponential convergence
of such algorithms, estimate explicitly the convergence rate, and
explore which properties of the stochastic input and the algorithm
affect the guaranteed convergence rate.
\end{abstract}



\pagestyle{myheadings}
\thispagestyle{plain}
\markboth{D. RACHINSKII AND M. RUDERMAN}{IDENTIFICATION OF PREISACH HYSTERESIS MODEL WITH STOCHASTIC INPUT}

\section{Introduction}

Preisach model \cite{9}, with its roots originated in magnetism
\cite{3,4}, is well-suitable for describing various
rate-independent hysteresis phenomena in a closed input-output
analytic form. The Preisach hysteresis operator
$y(k)=(\mathcal{H}[x])(k)$ models
an inpiut-output relationship with an erasable memory of previous
input states $x(j)$, where $j<k$. The input-output time
series $(x,y)(k)$ constitute monotonic piecewise continuous
hysteresis curves, see Figure 1 (right).

\begin{figure}[!h]
\centering
\includegraphics[width=0.7\columnwidth]{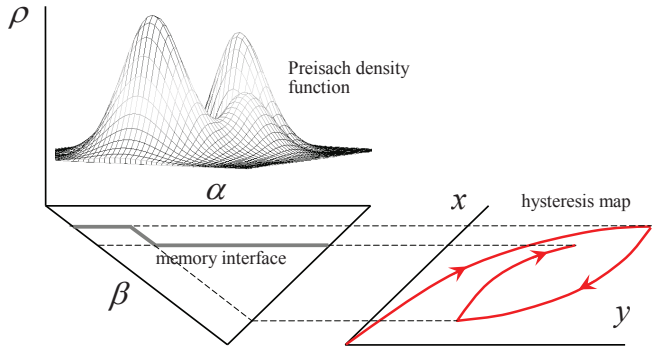}
\caption{Preisach density function $\rho(\alpha,\beta)$ over the plane
$(\alpha,\beta)$ (left) and input-output curves
$y=(\mathcal{H}[x])(t)$ of the Preisach operator (right).} \label{fig:0}
\end{figure}

According to the representation theorem of Mayergoyz,
the Preisach hysteresis model can be fitted to any set of input-output data
that possesses two properties: the so-called wiping-out of local input extremum values and congruency of any two hysteresis
loops created by the same two consecutive input extrema; see \cite{3} for more details of
the formalism and properties of the Preisach hysteresis model.

The use of models of hysteresis in magnetism, as well as
in other natural, engineering, and social sciences, confronts with
problems of identification of parameters, which requires to obtain a sufficient number of observations of
hysteresis system states  \cite{da4,da4'}. The
phenomenological nature of the Preisach model, which is based on
superposition of multiple elementary hysteresis operators
called non-ideal relays (switches), and the representation theorem of Mayergoyz allow to apply
the model to various hysteretic systems using a phenomenological argument or the black box approach
without complex analysis of such systems based on first principles \cite{da1,da2,da3,da4'',da5,da6,dimian,iyer}.
Importantly, the representation theorem provides the Preisach model with straightforward means
for identification of parameters from dedicated input-output data. Here it
should be recalled that the Preisach model is entirely
parameterized by the so-called Preisach density function $\rho=\rho(\alpha,\beta)$ defined
over the $(\alpha, \beta)$ plane, as schematically shown in Figure
1 (left). The $\alpha$ and $\beta$
coordinates, with $\beta\geq\alpha$, are the `on' and `off'
switching thresholds of a single relay operator, and the
$\rho$-function represents the weighted
$(\alpha,\beta)$-distribution of these relays over the possible threshold values.
Each relay switches from state $0$ (`off') to state $1$ (`on') and backwards in response to variations of the
same input time series called the input $x(k)$ of the Preisach model; the output time series $y(k)$ of the model is defined as the sum of states of all the relays.
The identification algorithm of Mayergoyz
requires to discretize the Preisach model and assumes a
sufficient number of measured first-order descending
(FOD) input-output curves \cite{3}. The higher the required accuracy of
hysteresis modeling is, the smaller the discrete step on the
$(\alpha, \beta)$ plane should be (cf. with Figure 1, left), and
the more accurate FOD curve measurements are required.
The volume of data, which should be first recorded before the
direct Mayergoyz identification method can be applied for calculation of the density function $\rho$,
rapidly increases with increasing target accuracy.

From this point of view, the motivation for this work is to explore the
possibilities of applying recursive/iterative identification
methods for which only the previous (or, a few recent) values of the
input-output data stream should be available at any instant.
Recursive (also known as on-line)
identification methods can be required when
dealing with time-varying processes, i.e., where the hysteresis
behavior changes (usually slowly) with time depending on
certain non-controllable operational or environmental conditions,
such as varying ambient temperature, force fields, temporal relaxation, etc.
The applicability of the direct Mayergoyz identification method
can also be limited by a strong uncontrollable noise capable
of corrupting the designed test input signal.
Furthermore, in
various engineering applications which use hysteresis models,
e.g., for process monitoring or control, a simplified
identification with no special (expert) knowledge as for design of
dedicated input processes (such as a process creating the FOD curves)
can be required. Here an algorithm which is capable of using data generated by a stochastic input processes (with certain
well-defined properties) may be advantageous.

Surprisingly, little number of recursive methods
suitable for identification of parameters of the Preisach hysteresis model has been
analyzed theoretically or experimentally. A
gradient-based recursive identification scheme has been proposed in
\cite{2} and experimentally evaluated on a magnetostrictive
actuator system which exhibits hysteresis between the relative
displacement and the actuator current. The error convergence for this scheme has been estimated for relatively rough discretization of
the $(\alpha, \beta)$ plane (up to 25 steps in the input domain).
Another direct recursive identification
method has been  proposed  in \cite{1}. This method updates
the values of the Preisach density function in the so-called switching region of the $(\alpha, \beta)$ plane,
which defines an output increment
in response to a random input increment
at each iteration step. This
method has been tested and evaluated numerically in the context of modeling
the hysteretic constitutive relationship between the magnetic field and magnetization
in a ferromagnetic material.
However no rigorous analysis of the convergence and its
properties, such as the convergence rate, has been done insofar.

In this paper, we prove the exponential convergence of the recursive
identification algorithm proposed in \cite{1}. We first show that the error of the algorithm
monotonically decreases (Section \ref{sec2})
and establish an estimate for the rate of convergence for any input time series (Section \ref{sec3}).
Our main result (Section \ref{sec4}) establishes that the error $E_k$
almost surely converges to zero exponentially, $
\|E_k\|\le C e^{-k\lambda} \|E_0\|$, with the rate satisfying an explicit estimate $\lambda\ge \mu(N_0, \langle T\rangle,\langle 1/T\rangle)>0$
if the input time series generates a regular Markov chain in the space of states of the Preisach model
(the latter condition is generic and is satisfied for a wide class of input Markov processes).
Here $N_0$ is a function of the number of nodes in the domain of the density function on the $(\alpha,\beta)$ plane,
$T$ is a certain stopping time associated with the input process, and $\langle T\rangle,\langle 1/T\rangle$ are the mean values
of $T$ and its inverse.
Essentially, $T$ is the length of the time interval during which fluctuations
of the input randomly create all the segments of all the FOD input-output curves prescribed by the classical
Mayergoyz identification algorithm.
If all this information was recorded, a complete identification of the density function
was achievable from the input-output data collected within $T$ time steps.
The recursive algorithm considered here is not capable of recovering the density function exactly as it
updates the approximation to the density function at each time step without recording
past input-output data. However, we guarantee that the error decreases
by a certain percentage over $T$ time steps, $\|E_{k+T}\|\le e^{- c/(N_0 T)}\|E_k\|$.
It is worth noting that the random time $T$ is a characteristic of the input process.
Overall, our estimates suggest that the smaller the mean value of $T$ is
and the less nodes the grid has (smaller $N_0$), the faster is the guaranteed exponential convergence of the recursive algorithm.
We test these conclusions by considering two examples of random input processes
and comparing their rates of convergence in Section \ref{sec5}.

\section{Monotonic decrease of the error}\label{sec2}
Let $\rho=\rho(p)$ be a density function of the Preisach operator where $p=(\alpha,\beta)$ denotes a point in the half-plane
$\Pi =\{p=(\alpha,\beta): \alpha\le\beta\}$. Each point $p$ represents a non-ideal
relay with the threshold values $\alpha$ and $\beta$.
Denote by $S_k=S(t_k)$ the subset of the half-plane $\Pi$ where relays are `on' (in state 1) at a moment $t_k=k\ge0$
(we normalize the time step to 1);
the other relays are `off' (in state 0).
Then the output $y_k=y(t_k)$ of the Preisach model at this moment equals
$$
y_k=\int_{S_k} \rho (p)\,dp
$$
where $dp=d\alpha d\beta$. The set $S_k$ is called the state of the Preisach model.
The state $S_k$ and the output $y_k$ change in response to variations of
the input $x_k=x(t_k)$ of the model as the non-ideal relays switch between states 0 and 1.
For discrete time input series, which is the case considered here,
$$
S_{k+1}=
\left\{
\begin{array}{cll}
S_k \cup \{(\alpha,\beta)\in \Pi: \beta\le x_{k+1}\} & {\rm if } & x_{k+1}> x_k\\~\\
S_k \setminus \{(\alpha,\beta)\in\Pi: \alpha\ge x_{k+1}\} & {\rm if } & x_{k+1}< x_k\\~\\
S_k& {\rm if } & x_{k+1}= x_k
\end{array}
\right.
$$
This rule defines the time series of the state and output, $S_k$ and $y_k$, for any given
input sequence $x_k$, $k\ge 0$ and a given initial state $S_0$ (see, for example, \cite{c1,c2,c3, ga1,ga2,ga3,ga4,ga5,new,new1}
for more details).

In the following identification algorithm we use
 an estimate $\hat \rho_k=\hat \rho_k(p)$, $p\in\Pi$ of the
density function, which is updated at each moment $t_k$. The
 estimate $\hat y_k=\hat y(t_k)$
of the output and the error $e_k=y_k-\hat y_k$ between the
observed and modeled output values equal, respectively,
$$
\hat y_k=\int_{S_k} \hat\rho_k(p)\,dp;\qquad e_k=\int_{S_k} (\rho(p)-\hat\rho_k(p))\,dp.
$$
Comparing states $S_k$ and $S_{k+1}$, denote $\Omega^+_k=S_{k+1}\setminus S_k$. This is the set of those relays that switch on during the time step from $t_k$ to
$t_k+1$.
Similarly, consider the set $\Omega^-_k=S_k\setminus S_{k+1}$ of relays that switch off during the same time step.
Then, over this
time step, the error between the observed output and the output modeled
with the density function $\hat\rho_k$ receives the increment
$$
\Delta e_k=\int_{\Omega^+_k} (\rho(p)-\hat\rho_k(p))\,dp - \int_{\Omega^-_k}(\rho(p)-\hat\rho_k(p))\,dp.
$$

Following \cite{1}, we update the estimate of the true density function $\rho$ according to the formula
\begin{equation}\label{scheme}
\hat \rho_{k+1}(p)=\hat \rho_k(p) + r_k(p) \quad {\rm with} \quad r_k(p)=
\left\{
\begin{array}{cll}
0 & {\rm if } & p\not\in \Omega_k=\Omega_k^+\cup \Omega_k^-\\~\\
\frac{\Delta e_k}{A_{\Omega_k}}  & {\rm if } & p\in \Omega_k^+\\~\\
-\frac{\Delta e_k}{A_{\Omega_k}}  & {\rm if } & p\in \Omega_k^-
\end{array}
\right.
\end{equation}
where $A_{\Omega_k}$ denotes the area of the switching region $\Omega_k$.

Let us first show that the mean square distance between the exact and estimated density is monotonically non-increasing with iterations.

\begin{lemma} For any initial estimate $\hat \rho_0$, the mean square distance
$$
\|\hat \rho_k -\rho\|=\left(\int_\Pi (\hat \rho_k (p)-\rho(p))^2\,dp \right)^{1/2}
$$
between $\hat\rho_k$ and $\rho$ is a non-increasing function of $k$.
\end{lemma}

{\em Proof}.
According to the update rule,
$$
\|\hat \rho_{k+1} -\rho\|^2 = \int_\Pi (\hat \rho_k (p)+r_k(p)-\rho(p))^2\,dp
$$
$$
=\int_\Pi (\hat \rho_k (p)-\rho(p))^2\,dp + \int_{\Omega_k} r^2_k(p)\,dp + 2 \int_{\Omega_k} r_k(p) (\hat \rho_k(p)-\rho(p))\,dp
$$
$$
=\|\hat \rho_{k} -\rho\|^2+ \frac{(\Delta e_k)^2}{A_{\Omega_k}} + \frac{2\Delta e_k}{A_{\Omega_k}}\left(
\int_{\Omega_k^+} (\hat \rho_k(p)-\rho(p))\,dp-\int_{\Omega_k^-} (\hat \rho_k(p)-\rho(p))\,dp
\right)
$$
$$
=\|\hat \rho_{k} -\rho\|^2+ \frac{(\Delta e_k)^2}{A_{\Omega_k}} - \frac{2(\Delta e_k)^2}{A_{\Omega_k}}=\|\hat \rho_{k} -\rho\|^2- \frac{(\Delta e_k)^2}{A_{\Omega_k}}.\qquad\endproof
$$

We see that
\begin{equation}\label{basic}
\|\hat \rho_{k+1} -\rho\|^2=\|\hat \rho_{k} -\rho\|^2- \frac{(\Delta e_k)^2}{A_{\Omega_k}},
\end{equation}
hence the mean square distance between the exact and estimated Preisach density functions not only never increases, but actually strictly decreases each time the output increment error $\Delta e$ is non-zero.
The aim of the following argument is to show that, for reasonable inputs, the error $\Delta e_k$ is non-zero ``often enough''
(as long as $\hat \rho_k\ne \rho$) to ensure
the exponential convergence $\hat \rho_k\to\rho$.

\section{Estimation of the rate of convergence}\label{sec3}
From now on, we will consider the discretized version of the Preisach model.
We will  assume that the input is a discrete Markov process $X_k$ with $N+1$ states $x_i=i\varepsilon$, where $\varepsilon=1/N$ and $i=0,\ldots,N$.
In this discrete setting, the set $\mathcal{S}$ of all possible states $S$ of the model is also finite.
Furthermore, we can assume without loss of generality that the density $\rho$ (as well as its approximations $\hat\rho_k$) is concentrated
at the finite set of nodes
\begin{equation}\label{mesh}
\{(\alpha,\beta)=((\ell-1/2)\varepsilon, (m-1/2)\varepsilon): 1\le \ell\le m\le N\}
\end{equation}
that form a uniform rectangular mesh within the right triangle
$$
\Lambda_{supp}=\{(\alpha,\beta): 0\le\alpha\le\beta\le1 \}
$$
of the half-plane $\Pi$;
these nodes will be denoted $p_i=(\alpha_i,\beta_i)$, where $i\in I$ with the total number of nodes $\# (I)=M=N(N+1)/2$.
In other words, we consider the density functions of the form $\rho(p)=\varepsilon^2 \sum_{i=1}^M \rho(p_i) \delta(p-p_i)$.
It is convenient to rescale the mesh step to unity by setting
$$
y_k=\sum_{\stackrel{i\in I}{p_i\in S_k}}\rho(p_i), \quad \hat y_k=\sum_{\stackrel{i\in I}{p_i\in S_k}}\hat \rho_k(p_i), \quad
\Delta e_k=\sum_{\stackrel{i\in I}{p_i\in \Omega_k^+}}(\rho(p_i)-\hat\rho_k(p_i))-\sum_{\stackrel{i\in I}{p_i\in \Omega_k^-}}(\rho(p_i)-\hat\rho_k(p_i))
$$
and, simultaneously, setting
$$
A_{\Omega_k}=N_k:= \#( i\in I: p_i\in \Omega_k)
$$
in the update rule \eqref{scheme} with $N_k$ denoting the number of the mesh nodes in the switching region $\Omega_k$.
We note that for this discrete model,
at each time step
either $\Omega_k^+=\emptyset$, $\Omega_k=\Omega_k^-$ or $\Omega_k^-=\emptyset$, $\Omega_k=\Omega_k^+$.
With this notation, formula \eqref{basic} can be written as
\begin{equation}\label{a}
\|E_{k+1}\|^2
=\|E_{k} \|^2- N_k (E_{k+1}(p)-E_k(p))^2, \qquad \forall p\in\Omega_k,
\end{equation}
where $E_k$ is the error
$$
E_k(p)=\hat \rho_k(p)-\rho(p)
$$
and $\|E_k\|=\left(\sum_{i\in I} E_k(p_i)^2\right)^{1/2}$.
Relationship \eqref{a} can be also written as
\begin{equation}\label{e}
\|E_{k+1}\|^2=\|E_{k} \|^2-\|E_{k}-E_{k+1}\|^2.
\end{equation}

We define the rate of convergence of the iterative scheme \eqref{scheme} as
\begin{equation}\label{lambda}
\lambda=-\limsup_{k\to\infty} \frac1k \ln\frac{\|E_k\|}{\|E_0\|}
=-\limsup_{k\to\infty}\frac1k \sum_{i=1}^k \ln\frac{\|E_i\|}{\|E_{i-1}\|}.
\end{equation}

In order to estimate the rate of convergence of the identification algorithm, we will use
a certain stopping time $T$ associated with the input process.
With a given realization of the input process $X_k$ we associate successive time intervals
$$
1\le k\le T_1,\ \ T_1+1\le k\le T_1+T_2,\ \ T_1+T_2+1\le k\le T_1+T_2+T_3,...
$$
During each time interval we require the process to create
a set of Preisach states satisfying the following properties.
We require that for every node $p_i=(\alpha_i,\beta_i)$
there would be two states $S(t_{k'})$ and $S(t_{k'}+1)$ achieved at two successive moments\footnote{It
is easy to extend the algorithm and results to the case when these moments are not necessarily successive.
We consider successive moments for simplicity.}
within the time interval $[T_1+\cdots+T_{k-1}+1,T_1+\cdots+T_k]$ such that
$$
(S({t_{k'}})\setminus S(t_{k'}+1))\cup
(S(t_{k'}+1)\setminus S(t_{k'}))=\Lambda_i
$$
where
\begin{equation}\label{defi}
\Lambda_i=\{(\alpha,\beta): \alpha_i-\varepsilon/2\le \alpha\le\beta\le \beta_i+\varepsilon/2\}.
\end{equation}
As soon as this requirement is satisfied for all the nodes $(\alpha_i,\beta_i)$, $i\in I$, of the mesh, the new time interval
$[T_1+\cdots+T_{k}+1,T_1+\cdots+T_{k+1}]$ starts. The set \eqref{defi}
is a right triangle with the vertex $(\alpha_i-\varepsilon/2,\beta_i+\varepsilon/2)$
and the hypothenuse on the line $\alpha=\beta$. Therefore, we will simply say
that the input creates all the possible triangles
on the half-plane $\Pi$ within each time interval $[T_1+\cdots+T_{k-1}+1,T_1+\cdots+T_k]$.

For example, this requirement will be satisfied if for every $(\alpha_i,\beta_i)$ there will
be three successive time moments $t_k=k$, $t_{k+1}=k+1$ and $t_{k+2}=k+2$ during the time interval $[T_1+\cdots+T_{k}+1,T_1+\cdots+T_{k+1}]$ such that
either
\begin{equation}\label{d1}
x(t_k)=\beta_j+\varepsilon/2,\ \ x(t_{k+1})=\alpha_i-\varepsilon/2,\ \ x(t_{k+2})=\beta_i+\varepsilon/2
\end{equation}
for some $\beta_j\ge\beta_i$, or
\begin{equation}\label{d2}
x(t_k)=\alpha_j-\varepsilon/2,\ \ x(t_{k+1})=\beta_i+\varepsilon/2,\ \ x(t_{k+2})=\alpha_i-\varepsilon/2
\end{equation}
for some $\alpha_j\le\alpha_i$.
A less restrictive requirement is that
the same holds but the moments $t_k$, $t_{k+1}$
are not necessarily successive and the input stays between $\alpha_i-\varepsilon/2$ and $\beta_j+\varepsilon/2$ for $t_k\le t\le t_{k+1}$ in the first case
(between $\alpha_j-\varepsilon/2$ and $\beta_i+\varepsilon/2$ for $t_k\le t\le t_{k+1}$ in the second case).

\begin{theorem}\label{th1}
The rate of convergence \eqref{lambda} of the iterative scheme \eqref{scheme} satisfies the estimate
\begin{equation}\label{est}
\lambda \ge \frac{1}{32N_0} \liminf_{k\to\infty} \frac1k \sum_{i=1}^{m(k)} \frac{1}{1+T_i}
\end{equation}
where $m=m(k)$ denotes the largest integer such that $k\ge T_1+\cdots+T_m$ and
\begin{equation}\label{n0}
N_0=\frac{N(N+1)^2}{4}+\frac{N^2(N^2-1)}{24}.
\end{equation}
\end{theorem}


In the next section, we will show that the limit in \eqref{est} is positive almost surely
and
can be found from the stationary distribution of the input process under natural assumptions.

{\em Proof of Theorem \ref{th1}}.
Define $\tau_0=0$, $\tau_m=T_1+\cdots+T_{m}$ and
assume that the input creates all the possible triangles during each time interval $[\tau_{k-1},\tau_k]$.
Set
$$
\sigma=\|E_{\tau_{m-1}}\|^2-\|E_{\tau_{m}}\|^2.
$$
According to formula \eqref{e},
$$
\sigma=\sum_{k=\tau_{m-1}}^{\tau_{m}-1} \|E_{k+1}-E_{k}\|^2.
$$
We split this sum into $M$ sums, where $M$ is the total number of nodes:
$$
\sigma=\sum_{i=1}^M \theta_i \sum_{k=\tau_{m-1}}^{\tau_{m}-1} \|E_{k+1}-E_{k}\|^2
$$
where the coefficients $\theta_i$ satisfy
$$
\theta_1 + \cdots + \theta_M=1.
$$
Now, we replace the upper limit $\tau_{m}-1$ in the internal sum by the
moment $t_i\in [\tau_{m-1},\tau_{m}-1]$ when the switching area during the time step from $t_i$ to $t_{i}+1$ is the triangle
$\Lambda_i=\{(\alpha,\beta): \alpha_i-\varepsilon/2\le \alpha \le \beta \le \beta_i+\varepsilon/2\}$. In this way, we obtain the estimate
$$
\sigma\ge \sum_{i=1}^M \theta_i \sum_{k=\tau_{m-1}}^{t_{i}} \|E_{k+1}-E_{k}\|^2.
$$
At the time step from $t_i$ to $t_{i}+1$ we have
$$
\|E_{t_i+1}-E_{t_i}\|^2=\frac{(A_i^{t_i})^2}{N_i}
$$
where $A_i^t=\sum_{p\in\Lambda_i} E_t(p)$ denotes the sum of $E_{t}(p)$ over the triangle $\Lambda_i$ at the moment $t$
and $N_i$ is the number of nodes in the triangle. Hence,
$$
\sigma\ge \sum_{i=1}^M \theta_i \frac{(A_i^{t_i})^2}{N_i} + \sum_{i=1}^M \theta_i \sum_{k=\tau_{m-1}}^{t_{i}-1} \|E_{k+1}-E_{k}\|^2.
$$
Using the inequality
$$
\sum_{k=\tau_{m-1}}^{t_{i}-1} \|E_{k+1}-E_{k}\|^2\ge \frac{1}{t_i-\tau_{m-1}}\left \|\sum_{k=\tau_{m-1}}^{t_{i}-1} (E_{k+1}-E_{k})\right\|^2,
$$
we obtain
$$
\sum_{k=\tau_{m-1}}^{t_{i}-1} \|E_{k+1}-E_{k}\|^2\ge \frac{\|E_{t_i}-E_{\tau_{m-1}}\|^2}{t_i-\tau_{m-1}}\ge \frac{\|E_{t_i}-E_{\tau_{m-1}}\|^2}{T_m}
$$
and therefore
$$
\sigma\ge \sum_{i=1}^M \theta_i \left(\frac{(A_i^{t_i})^2}{N_i} + \frac{\|E_{t_i}-E_{\tau_{m-1}}\|^2}{T_m}\right).
$$

Let us consider one term in this sum:
\begin{equation}\label{xx}
\frac{(A_i^{t_i})^2}{N_i} + \frac{\|E_{t_i}-E_{\tau_{m-1}}\|^2}{T_m}\ge
\frac{1}{N_i}
\left(\sum_{p\in \Lambda_i} {E_{t_i}(p)}\right)^2 +
\frac{1}{T_m} \sum_{p\in\Lambda_i} (E_{t_i}(p)-E_{\tau_{m-1}}(p))^2.
\end{equation}
The right hand side of this inequality can be rewritten as
$$
\frac{1}{N_i}
\left(\sum_{p\in \Lambda_i} ({E_{\tau_{m-1}}(p)+\chi_p})\right)^2 +
\frac{1}{T_m} \sum_{p\in\Lambda_i} \chi_p^2
$$
where $\chi_p=E_{t_i}(p)-E_{\tau_{m-1}}(p)$. Considering the minimization problem
$$
V(\xi_1,\ldots,\xi_{N_i})=\frac{1}{N_i}
\left(\sum_{p\in \Lambda_i} ({E_{\tau_{m-1}}(p)+\xi_p})\right)^2 +
\frac{1}{T_m} \sum_{p\in\Lambda_i} \xi_p^2 \to {\rm min}
$$
with respect to the variables $\xi_p$, we obtain the equations
$$
\frac{1}{N_i}
\left(\sum_{p\in \Lambda_i} ({E_{\tau_{m-1}}(p)+\xi_p})\right) +
\frac{\xi_p}{T_m}=0, \qquad p\in\Lambda_i,
$$
for the extremum, hence the minimum is achieved at the point
$$
\xi_1=\cdots=\xi_{N_i}=-\frac{T_m}{N_i(1+T_m)}\sum_{p\in\Lambda_i} E_{\tau_{m-1}}(p)
$$
and the global minimum value equals
$$
V_{min}=\frac{1}{N_i(1+T_m)}\left(\sum_{p\in\Lambda_i} E_{\tau_{m-1}}(p)\right)^2=\frac{(A_{i}^{\tau_{m-1}})^2}{N_i(1+T_m)}.
$$
Thus, this expression is the lower estimate for the the right hand side of inequality \eqref{xx}
and we conclude that
$$
\sigma\ge \frac{1}{1+T_m} \sum_{i=1}^M \frac{\theta_i (A_{i}^{\tau_{m-1}})^2}{N_i}.
$$
We rearrange the sum in the right hand side as follows:
$$
\sigma\ge \frac{1}{4(1+T_m)}\sum_{i=1}^M \left( \frac{\theta_i (A_{i}^{\tau_{m-1}})^2}{N_i}+
\frac{\theta_j (A_{j}^{\tau_{m-1}})^2}{N_j}+\frac{\theta_k (A_{k}^{\tau_{m-1}})^2}{N_k}+\frac{\theta_\ell (A_{\ell}^{\tau_{m-1}})^2}{N_\ell} \right)
$$
where $j=j(i), k=k(i), \ell=\ell(i)$ are defined so that
$\Lambda_j=\{(\alpha,\beta): \alpha_i+\varepsilon/2\le \alpha \le \beta \le \beta_i+\varepsilon/2\}$,
$\Lambda_k=\{(\alpha,\beta): \alpha_i-\varepsilon/2\le \alpha \le \beta \le \beta_i-\varepsilon/2\}$,
$\Lambda_\ell =\{(\alpha,\beta): \alpha_i+\varepsilon/2\le \alpha \le \beta \le \beta_i-\varepsilon/2\}$
(for the nodes $(\alpha_i,\beta_i)$ on the diagonal $\beta=\alpha$ we set $A_j=A_k=A_\ell=0$).

We choose
$$
\theta_i=\frac{N_i}{N_0}
$$
with
$$
N_0=\sum_{i=1}^M N_i=\frac12 \sum_{k=1}^N k(k+1)(N-k+1)=\frac12\sum_{k=1}^N ((N+1)k + N k^2 -k^3)
$$
$$
=\frac{N(N+1)^2}{4}+\frac{N^2(N+1)(2N+1)}{12}-\frac{N^2(N+1)^2}{8}=\frac{N(N+1)^2}{4}+\frac{N^2(N^2-1)}{24}
$$
where the total number of nodes is $M=N(N+1)/2$. With this choice of $\theta_i$,
$$
\sigma\ge \frac{1}{4(1+T_m)N_0}\sum_{i=1}^M \left( (A_{i}^{\tau_{m-1}})^2+
(A_{j}^{\tau_{m-1}})^2+(A_{k}^{\tau_{m-1}})^2+(A_{\ell}^{\tau_{m-1}})^2\right).
$$
We now note that for each node $i$
$$
A_{i}^{\tau_{m-1}}-
A_{j}^{\tau_{m-1}}-A_{k}^{\tau_{m-1}}+A_{\ell}^{\tau_{m-1}}=E_{\tau_{m-1}}(i).
$$
Minimizing the function $W(\xi_1,\xi_2,\xi_3,\xi_4)=\xi_1^2+\xi_2^2+\xi_3^2+\xi_4^2$ under the constraint
$\xi_1+\xi_2+\xi_3+\xi_4=E_{\tau_{m-1}}(i)$, we see that $\min W=(E_{\tau_{m-1}}(i))^2/4$, hence
$$
\sigma=\|E_{\tau_{m-1}}\|^2-\|E_{\tau_{m}}\|^2\ge \frac{1}{4(1+T_m)N_0}\sum_{i=1}^M \frac{(E_{\tau_{m-1}}(i))^2}{4}=\frac{\|E_{\tau_{m-1}}\|^2}{16(1+T_m)N_0}.
$$
That is,
$$
\|E_{\tau_{m}}\|^2\le \left(1-\frac{1}{16(1+T_m)N_0}\right)\|E_{\tau_{m-1}}\|^2.
$$
As $\ln (1-x)\le -x$ for $0< x<1$, this implies
$$
-\ln \|E_{\tau_{m}}\|+ \ln \|E_{\tau_{m-1}}\|\ge \frac{1}{32(1+T_m)N_0}.
$$
Summing, we obtain
$$
-\ln \|E_{k}\|+ \ln \|E_{0}\|\ge \frac{1}{32N_0}\sum_{i=1}^m \frac{1}{1+T_i}
$$
where we use the largest $m=m(k)$ such that $k\ge T_1+\cdots+T_m$. Therefore,
the rate of convergence \eqref{lambda} of the iterative scheme satisfies \eqref{est}. \hfill \endproof

\section{Main result}\label{sec4}
As  the input increment $X_{k+1}-X_k$ and the state $S_k$ of the Preisach model
at a moment $k$ define the state $S_{k+1}$ at the next moment,
the input Markov chain $X_k$ defines the Markov chain $S_k$
in the state space $\mathcal{S}$ of the Preisach model describing the random evolution of the state.

Now, let us consider two more auxiliary Markov chains defined by the input process.
One is $Y_k=S_{T_1+\cdots+T_k}$. That is, any realization of the
Markov chain $S_k$ generates
 a realization of the auxiliary process $Y_{k}=S_{\tau_k}$ where all the triangles \eqref{defi}
 are created between any two moments $\tau_{k-1}=T_1+\cdots+T_{k-1}$ and $\tau_k=T_1+\cdots+T_k$\footnote{According to the definition
 of the times $\tau_k$, each moment $\tau_k$ is minimal in the sense that at least one triangle is not created between the moments $\tau_{k-1}$ and $\tau_k-1$.}.
 The finite state space of the Markov chain $Y_k$ is the space $\mathcal S$ of states of the Preisach model.
 Here and henceforth we assume that the Markov chain $S_k$ almost surely creates all the triangles \eqref{defi} in finite time,
 hence the Markov chain $Y_k=S_{\tau_k}$ is well defined.

We will assume that properties of the input Markov chain $X_k$
ensure that the Markov chain $Y_k=S_{\tau_k}$ is regular (ergodic), that is
some power of the transition probability matrix of the process $Y_k$ is strictly positive.

Secondly, we consider an auxiliary Markov chain $Z_k=(S_{T_1+\cdots+T_k},T_{k})$ with a countable number of states.
A pair $(\hat S, \hat T)$ belongs to the state space $\mathcal{Z}$ of this Markov chain if and only if
there is a Preisach state $\tilde S$ such that $P(S_{T_1}=\hat S, T_1=\hat T | S_0=\tilde S)>0$.
Let us observe that the transition probability $P(Z_{k+1}=Z' | Z_k=Z)$ from a state $Z=(S, \tau)\in \mathcal{Z}$
to a state $Z '= (S', \tau')\in \mathcal{Z}$ of the Markov chain $Z_k$ is, by its definition,
independent of the component $\tau$ of the initial state:
\begin{equation}\label{ptran}
P( S_{T_1+\cdots+T_{k+1}}=S',\ T_{k+1}=\tau' |\ S_{T_1+\cdots+T_k}= S,\ T_k=\tau)=: p_{S, S',\tau'}.
\end{equation}
As $Y_k=S_{\tau_k}$ is a regular Markov chain,
for every Preisach state $S\in \mathcal{S}$ there is at least one positive integer $T$ such that $(S,T)\in \mathcal Z$.


\begin{lemma}\label{l2}
Assume that the Markov chain $Y_k=S_{T_1+\cdots+T_k}$ is regular.
Then, the Markov chain $Z_k=(S_{T_1+\cdots+T_k},T_{k})$ is (a) irreducible; (b) positive recurrent; and, (c) aperiodic.
\end{lemma}

{\em Proof}. 

{\bf Irreducibility. } Consider any pair of states $\bar Z=(\bar S, \bar T), \hat Z=(\hat S, \hat T) \in \mathcal{Z}$.
By definition of the state space $\mathcal{Z}$, there is at least one state $\tilde Z=(\tilde S,\tilde T)\in \mathcal{Z}$
such that $P(S_{T_1}=\hat S, T_1=\hat T | S_0=\tilde S)>0$. On the other hand, as the process $Y_k=S_{\tau_k}$ is regular,
there is a positive integer $k_0$ such that $P(S_{T_1+\cdots+T_{k_0}}=\tilde S | S_0=\bar S)>0$.
Combining these two estimates and using the time invariance of the Markov chain $S_k$,
we obtain $P(S_{T_1+\cdots+T_{k_0+1}}=\hat S, T_{k_0+1}=\hat T | S_0=\bar S)>0$.
But this is the probability that the process $Z_k$ reaches the state $\hat Z$ from the state $\bar Z$ in $k_0+1$ steps.
As the probability is positive, we conclude that $Z_k$ is irreducible.

{\bf Positive recurrence. }
For any state $\hat Z=(\hat S, \hat T) \in \mathcal{Z}$ consider
the state $\tilde Z=(\tilde S,\tilde T)\in \mathcal{Z}$
such that $\hat p:=P(S_{T_1}=\hat S, T_1=\hat T | S_0=\tilde S)>0$.
Consider again a sufficiently large integer $k_0$ such that
the regularity of the process  $Y_k=S_{\tau_k}$ implies
$p_m:=\min \{P(S_{T_1+\cdots+T_{k_0}}=\tilde S | S_0=\bar S): \bar S\in \mathcal{S}\}>0$.
We see that the probability that the process $Z_k$ starting from the state $\hat Z$
returns to this state after $k_0+1$ time steps,
$P(S_{T_1+\cdots+T_{k_0+1}}=\hat S, T_{k_0+1}=\hat T | S_0=\hat S)$,
is greater or equal than $p_m\hat p>0$, hence the Markov chain $Z_k$ is recurrent.
Furthermore, $P(S_{T_1+\cdots+T_{k_0+1}}=\hat S, T_{k_0+1}=\hat T | S_0=\bar S)\ge p_m\hat p$
for any $\bar S\in \mathcal{S}$. Therefore, the probability that the first return time to the state $\hat Z$
exceeds $n(k_0+1)$ is less or equal than $(1-p_m\hat p)^n$ for any integer $n$.
Hence, the expected value of this first return time does not exceed the value
$$
\sum_{k=1}^{k_0+1} k + (1-p_m\hat p)\sum_{k=k_0+2}^{2k_0+2} k+ (1-p_m\hat p)^2 \sum_{k=2k_0+3}^{3k_0+3}k+\cdots
$$
As this sum is finite, the expected value of the first return time is finite, hence the process $Z_k$
is positive recurrent.

{\bf Aperiodicity. }
Using the same notation as above,
$\hat p=P(S_{T_1}=\hat S, T_1=\hat T | S_0=\tilde S)>0$.
From the regularity of the process $Y_k=S_{\tau_k}$ it follows
that $p(k):= P(S_{T_1+\cdots+T_{k}}=\tilde S | S_0=\hat S)>0$ for any
$k\ge k_0$. Therefore,
$P(S_{T_1+\cdots+T_{k+1}}=\hat S, T_1=\hat T | S_0=\hat S)>0$, that is the process
$Z_k$ returns to the state $\hat Z=(\hat S, \hat T)$ with positive probability
after $k$ time steps for any $k\ge k_0+1$. Hence the state $\hat Z$ is aperiodic.
Since this is true for any $\hat Z\in \mathcal Z$, the Markov chain $Z_k$ is aperiodic.
{}\hfill \endproof 

Lemma \ref{l2} ensures that
the finite Markov chain $Y_k$
has a strictly positive and unique stationary distribution $\pi_Y$.
Furthermore, the Markov chain $Z_k$ also
has a unique stationary distribution $\pi_{Z}$ and the ergodicity property
$$
\frac1k \sum_{i=1}^k f(Z_k) \stackrel{a.s.}{\rightharpoondown} E_{\pi_Z}[ f(Z)]
$$
holds for any initial state $Z_1$ and any function $f$ with a finite expected value
$$
E_{\pi_Z}[ f(Z)]<\infty
$$
with respect to the distribution $\pi_{Z}$ (see, for example, \cite{6}).
In particular, using $f(Z)=T$ where $Z=(S,T)$, we obtain
\begin{equation}\label{1T1}
\frac1k \sum_{i=1}^k T_k \stackrel{a.s.}{\rightharpoondown} E_{\pi_Z}[ T_k]=: T_*
\end{equation}
(the condition $T_*<\infty$ will be established in the proof of Theorem \ref{th2} below).
That is, the time average of the time intervals $T_k$ converges to the mean $T_*$ of $T_k$ with respect to the stationary
distribution of the chain $Z_k$. Similarly,
\begin{equation}\label{1T}
\frac1k \sum_{i=1}^k \frac{1}{1+T_k} \stackrel{a.s.}{\rightharpoondown} E_{\pi_Z}\left[\frac{1}{1+T_k}\right]=: T_{**}.
\end{equation}

Let us remark that the mean value $T_*$ is defined by
$$
T_*=\sum_{(S',\tau')\in \mathcal{Z}} \tau' \cdot (\pi_{Z})_{S',\tau'}
$$
where $(\pi_{Z})_{S',\tau'}$ is the probability to find the process $Z_k$ in the state $(S_{k},T_k)=(S',\tau')$ according to the stationary
distribution of this process.
According to \eqref{ptran}, for the stationary distribution,
\begin{equation}\label{nnn}
(\pi_{Z})_{S',\tau'}=\sum_{(S,\tau)\in \mathcal{Z}} (\pi_{Z})_{S,\tau}\cdot p_{S,S',\tau'}=\sum_{S\in \mathcal{S}} \Pi_{S}\cdot p_{S,S',\tau'}
\end{equation}
where $\Pi_S=\sum_{\tau}(\pi_{Z})_{S,\tau}$ and we sum over all the pairs $(S,\tau)\in\mathcal{Z}$
with the fixed $S$.
Summing \eqref{nnn} with respect to $\tau'$ over all pairs $(S',\tau')\in\mathcal{Z}$
with the fixed $S'$, we obtain
$$
\Pi_{S'}=\sum_{S\in \mathcal{S}} P_{S,S'} \Pi_S
$$
where $P_{S,S'}=\sum_{\tau'} p_{S,S',\tau'}$. As the stationary distribution $\pi_Y$ of the Markov chain
$Y_k=X_{T_1+\cdots+T_k}$ satisfies the same equation with the same transition probabilities $P_{S,S'}$ and the same normalization condition,
the uniqueness implies $\Pi=\pi_Y$. Hence, \eqref{nnn} is equivalent to
$$
(\pi_{Z})_{S',\tau'}=\sum_{S\in\mathcal{S}} (\pi_{Y})_S\cdot p_{S,S',\tau'}
$$
and, furthermore, the mean $T_*=E_{\pi_Z}[ T_k]$ of the time $T_k$ satisfies
\begin{equation}\label{mt}
E_{\pi_Z}[ T_k]=
\sum_{S\in\mathcal{S}} (\pi_{Y})_S \sum_{(S',\tau')\in \mathcal{Z}} \tau' \cdot  p_{S,S',\tau'}.
\end{equation}
That is, we average the time required to create all the triangles \eqref{defi} (that is, the transition time
from $Y_k$ to $Y_{k+1}$)
over the stationary distribution $\pi_Y$ of the initial state $Y_k=S$ (regardless of the destination state $Y_{k+1}=S'$).
Similarly,
\begin{equation}\label{mt1}
E_{\pi_Z}\left[ \frac{1}{1+T_k}\right]=
\sum_{S\in\mathcal{S}} (\pi_{Y})_S \sum_{(S',\tau')\in \mathcal{Z}} \frac{p_{S,S',\tau'}}{1+\tau'}.
\end{equation}

Now, we are ready to formulate the main result.

\begin{theorem}\label{th2}
Assume that the Markov chain $Y_k=S_{T_1+\cdots+T_k}$ induced by the input process $X_k$
in the Preisach state space $\mathcal{S}$ is regular.
Then the error of the iterative scheme \eqref{scheme} exponentially decreases
and the rate of the exponential convergence \eqref{lambda} satisfies the estimate
\begin{equation}\label{1'}
\lambda\ge \frac{ E_{\pi_Z}\left[ \frac{1}{1+T_k}\right]}{32N_0 \cdot E_{\pi_Z}[ T_k]}>0
\end{equation}
where $N_0$ is given by \eqref{n0} and $E_{\pi_Z}[ T_k]$, $E_{\pi_Z}\left[ \frac{1}{1+T_k}\right]$ are defined by \eqref{mt}, \eqref{mt1}.
\end{theorem}

According to \eqref{1'}, the lower bound for the rate of the exponential convergence
is controlled by the average time that the input process takes to create all the triangles
on the half-plane $\Pi$. One can assume that the shorter is this time, the faster is the
convergence. We will test this conjecture with numerical examples in the next section.
Also, the lower bound  decreases as $O(N^{-4})$ with the increasing density of the mesh on the half-plane $\Pi$.

{\em Proof of Theorem \ref{th2}}.
From $T_k\ge 1$ it follows that $E_{\pi_Z}\left[\frac{1}{1+T_k}\right]<\infty$, hence Lemma \ref{l2} implies relation \eqref{1T} \cite{6}.

Let us show that also $E_{\pi_Z}\left[{T_k}\right]<\infty$. Indeed,
we assumed that the Markov chain $S_k$ starting from any initial state $S_0$ almost surely creates all the triangles \eqref{defi}
in finite time. Therefore, there is an integer $\tau_*$ such that
the process $S_k$ creates all the triangles \eqref{defi} within the first $\tau_*$ time steps
with a positive probability $p_{S_0}\ge p_*>0$ for any $S_0\in\mathcal {S}$.
Hence, the probability \eqref{ptran} satisfies $p_{S,S',\tau'} \le (1-p_*)^n$
for every $\tau'\ge n\tau_*$ and every $S,S'\in\mathcal{S}$
and consequently formula \eqref{mt} implies $E_{\pi_Z}\left[{T_k}\right]<\infty$.

As the mean $E_{\pi_Z}\left[{T_k}\right]$ of the times $T_k$ is finite, using Lemma \ref{l2} we conclude that
\eqref{1T1} is valid.

Finally, let us see that
\begin{equation}\label{one}
\frac1k \sum_{i=1}^{m(k)} T_i \stackrel{a.s.}{\rightharpoondown} 1
\end{equation}
where $m=m(k)$ denotes the largest integer such that $k\ge T_1+\cdots+T_m$.
For any $\varepsilon>0$, the estimate
\begin{equation}\label{limin}
\liminf_{k\to\infty}\frac1k\sum_{i=1}^{m(k)} T_i<1-\varepsilon
\end{equation}
implies the existence of a subsequence $T_{i_n}$ satisfying
\begin{equation}\label{bn}
T_1+\cdots+T_{i_n}\le  \frac{T_{i_n+1}}{\varepsilon}.
\end{equation}
Since $p_{S,S',\tau'} \le (1-p_*)^{\tau'/\tau_*-1}$
for every positive integer $\tau'$ and every $S,S'\in\mathcal{S}$, the probability that
a sequence $T_i$ contains at least one element $T_{i_*+1}=\tau'$ and, simultaneously, satisfies
$\varepsilon(T_1+\cdots+T_{i_*})\le T_{i_*+1}=\tau'$ can be estimated as follows:
$$
P(T_{i+1}=\tau', \,  T_1+\cdots+T_{i}\le T_{i+1}/\varepsilon \ \ {\rm for\ at\ least\ one}\,\ i)
\le \frac{\tau'}{\varepsilon}(1-p_*)^{\tau'/\tau_*-1}.
$$
Therefore,
$$
P(T_{i+1}\ge \tau', \,  T_1+\cdots+T_{i}\le T_{i+1}/\varepsilon \ \ {\rm for\ at\ least\ one}\,\ i)
\le\sum_{j=\tau'}^\infty \frac{j}{\varepsilon}(1-p_*)^{j/\tau_*-1}.
$$
As the right hand side of this expression tends to zero with $\tau'\to\infty$ and $T_{i_n}\to\infty$ in \eqref{bn}, we conclude that relationship \eqref{limin}
holds with zero probability for each $\varepsilon>0$. This proves \eqref{one}.

Now, estimate \eqref{est} can be rewritten as
$$
\lambda\ge \frac{1}{32N_0} \liminf_{k\to\infty} \left(\frac1k \sum_{i=1}^{m(k)} T_i \right)  \left(\frac1{m(k)} \sum_{i=1}^{m(k)} T_i \right)^{-1}
 \left(\frac1{m(k)} \sum_{i=1}^{m(k)} \frac{1}{1+T_i}\right).
$$
Combining this formula with relations \eqref{1T1}, \eqref{1T} and \eqref{one}, and taking into account that $m(k)\to\infty$ as $k\to \infty$ almost surely
(because the Markov chain $S_{T_1+\cdots+T_k}$ is, by assumption, well defined), we obtain
\eqref{1'}.
\hfill \endproof 

\section{Numerical examples}\label{sec5}
%
%
%

In this section, we consider two examples of the input process and evaluate the convergence rate of the norm $\|E_{k}\|$
of the error for the iterative scheme \eqref{scheme}, and that depending on the quantities $T_{*}$ and
$N_{0}$ as in \eqref{n0} and \eqref{1T1}, respectively. These are the two factors that affect the convergence rate guaranteed by
Theorem \ref{th2}. The following numerical computations were performed
using the discrete dynamic Preisach (DDP) modeling
algorithm which is suitable for realization of the iterative scheme \eqref{scheme}. The DDP algorithm
has been introduced in \cite{7} and applied for inverse
hysteresis control in \cite{8}.

As the density function $\rho(\alpha,\beta)$ of the Preisach model, 
we used the
$\delta$-finction $\rho=\delta(p-p_0)$  located at the point $p_0=(\alpha,\beta)=(N\varepsilon/2, N\varepsilon/2)$.
The convergence in this case is relatively slow, because the algorithm makes a nonzero correction
to the approximating density $\hat \rho$ only when the switching region contains the point $p_0$.
Moreover, the correction is each time spread over multiple (at least $(N-1)/2$) nodes.
The initial approximation of the scheme was the uniform density $\hat \rho_0(\alpha,\beta)=a$ with a small positive $a$.

%
%
%

Two meshes with
$N^{\mathrm{I}}=31$ and $N^{\mathrm{II}}=101$ were considered (the odd number is
conditioned by the numerical implementation of DDP algorithm). The number of nodes
for these meshes,
$M^{\mathrm{I}}=496$ and $M^{\mathrm{II}}=5151$, differs by one order of magnitude.

\subsection{Stochastic inputs} \label{sub:2}

We will consider two types of the input Markov chain $X_{k}$.
%
%
The first input denoted by $X^{\mathrm A}_k$ is the Bernoulli (memoryless) process:
$$
P\bigl(X_{k}^{\mathrm{A}}= i \varepsilon \bigr)=\frac{1}{N+1}, \qquad i =0,...,N.
$$
The second input process denoted by $X^{\mathrm B}_k$ is
a modification of the Bernoulli process where the sign of the input increment alternates at each time step:
$$
P\bigl(X_{k+1}^{\mathrm{B}}= j \varepsilon | X_k^{\mathrm{B}}=i\varepsilon \bigr)=\left\{
\begin{array}{cl}
0, & j<i\\
\frac{1}{N+1-i}, & j \ge i
\end{array}
\right.
\qquad {\rm for\ odd}\ k;
$$
$$
P\bigl(X_{k+1}^{\mathrm{B}}= j \varepsilon | X_k^{\mathrm{B}}=i\varepsilon \bigr)=\left\{
\begin{array}{cl}
\frac{1}{i+1}, & j\le i\\
0, & j > i
\end{array}
\right.
\qquad {\rm for\ even}\ k,
$$
where $i,j=0,\ldots,N$.
Thus, a specific feature of this
process is that it produces continuously oscillating input
sequences.
Note that both processes A and B have no knowledge of dynamics
in the state space $\mathcal{S}$ of the Preisach model.

We now show that the above two stochastic input  processes satisfy the conditions of Theorem \ref{th2}.

\begin{lemma}\label{l3}
The Markov chain $Y_k=S_{T_1+\cdots+T_k}$ induced by each of the processes A and B
in the Preisach state space $\mathcal{S}$ is regular.
\end{lemma}

{\em Proof}.
A state $S$ of the Preisach model will be identified with a polyline connecting
the point $(\alpha,\beta)=(0,N\varepsilon)$ with the line $\alpha=\beta$
on the half-plane $\Pi$. This polyline consists of vertical and horizontal links
where the length of each link is a multiple of $\varepsilon$.
The relays are in state 1 below the polyline and in state 0 above the polyline \cite{3}.

The following finite sequence of input values will be called a {\em standard sequence}:
\begin{equation}\label{gg}
\begin{array}{llllllllllllllllll}
0, & x_1,\\
0, & x_2, &x_1, & 0, & x_2, \\
0, &x_3, &x_1, & 0, & x_3, &x_2, & 0, & x_3, \\
0, & x_4, & x_1, & 0, &x_4, &x_2, & 0, & x_4, &x_3, &  0, & x_4,  \\
\vdots
\\
0, &x^N, &x^1, & 0, & x_N, &x_2, & 0, & x_N, & x_3, & \cdots, &0, &x^N, &x^{N-1}, &    0, &x^N, \\
0,
\end{array}
\end{equation}
where $x_i=i \varepsilon$. This input sequence produces all the possible triangles \eqref{defi}.
By definition, each of the processes A and B has a positive
transition probability from any state $x_i$ to any state $x_j$.
Therefore,  for any given initial state, the input makes the {\em standard sequence}
of steps and creates all the triangles with a positive probability.
Hence, the process $Y_k=S_{T_1+\cdots+T_k}$ in the Preisach state space is well defined
(that is the times $T_i$ are almost surely all finite).

Now, it suffices to show that, given any pair of states $S', S''\in {\mathcal S}$, the transition probability
of the Markov chain $Y_k=S_{T_1+\cdots+T_k}$  from $S'$ to $S''$ is positive.
Denote by $P_0',\ldots,P_{K'}'$ the corners of the polyline $S'$
and by $P_0'',\ldots,P_{K''}''$ the corners of the polyline $S''$,
where $P_0'=P_0''=(0,\varepsilon N)$, $\alpha_{K'}'=\beta_{K'}'$, $\alpha_{K''}''=\beta_{K''}''$
and we use the notation $P_i'=(\alpha'_i,\beta_i')$, $P_i''=(\alpha''_i,\beta_i'')$.
For any state $S''=P_0'' P_1''\cdots P_{K''}''\in {\mathcal S}$ (except for the state
$S''=\{(\alpha,\beta): \alpha=0, \beta\in[0,N\varepsilon]\}$ consisting of one vertical link),
we define an input sequence {\em associated} with $S'$; this sequence is given by:
\begin{itemize}
\item
$\beta_2'',\alpha_3'',\beta_4'',\alpha_5'',\ldots, \alpha_{K''-1}'',\beta_{K''}''$ if the link
$P_0''P_1''$ is vertical and the link $P_{K''-1}''P_{K''}''$ is horizontal;

\item
$\beta_2'',\alpha_3'',\beta_4'',\alpha_5'',\ldots, \beta_{K''-1}'',\alpha_{K''}''$ if both the links
$P_0''P_1''$ and $P_{K''-1}''P_{K''}''$ are vertical;

\item
$\beta_1'',\alpha_2'',\beta_3'',\alpha_4'',\ldots, \alpha_{K''-1},\beta_{K''}''$ if both the links
$P_0''P_1''$ and $P_{K''-1}''P_{K''}''$ are  horizontal;

\item
$\beta_1'',\alpha_2'',\beta_3'',\alpha_4'',\ldots, \beta_{K''-1},\alpha_{K''}''$ if the link
$P_0''P_1''$ is horizontal and the link $P_{K''-1}''P_{K''}''$ is vertical.
\end{itemize}
The input sequence associated with the state $S''$ induces the transition
of the process $S_k$ from the state consisting of one vertical link to the state $S''$.

Depending on a few features of the states $S',S''$, we will define explicitly
a finite sequence of input values, which induces the transition
of the Markov chain $Y_k$ from the state $Y_k=S'$
to the state $Y_{k+1}=S''$; the probability of such transition
is positive, because the input sequence is finite.
We call such an input sequence a {\em transition input sequence}.

First, consider any state $S'=P_0'P_1'\cdots P_{K'}'$ which has a vertical link $P_{K'-1}' P_{K'}'$.
Consider the modified standard sequence \eqref{gg} where we replace every instance of $0$, except for the first and the last one,
with the pair of entries $0, x_1$:
\begin{equation}\label{ggg}
\begin{array}{lllllllllllllllllllllll}
0, & x_1,\\
 0, & x_1, & x_2, &x_1, &0, & x_1, & x_2, \\
0, & x_1, &x_3, &x_1, & 0, & x_1, & x_3, & x_2, &   0, & x_1, & x_3,   \\
\vdots
\\
0, & x_1, &x^N, &x^1,  &0, & x_1, &x^N, &x^{2}, &  0, & x_1, &x^N, & x_3, & \cdots, & 0, & x_1, &x^N,\\
0.
\end{array}
\end{equation}

\begin{itemize}
\item
The input sequence \eqref{ggg} induces the transition of the Markov chain $Y_k=S_{T_1+\cdots+T_k}$
from the state $S'$ to the state $S''=\{(\alpha,\beta): \alpha=0, \beta\in[0,N\varepsilon]\}$ consisting of one vertical link.

\item
If the polyline $S''=P_0''P_1''\cdots P_{K''}''$ has more than one link and its link $P_{K''-1}'' P_{K''}''$ is vertical, then
a transition input sequence from $S'$ to $S''$ is the sequence \eqref{ggg}, where we omit the successive four entries $x_1, \beta''_{K''-1}, \beta_{K''}'',0$,
which is followed by the input sequence associated with the state $S''$.

\item If the polyline $S''=P_0''P_1''\cdots P_{K''}''$ has more than two links and its link $P_{K''-1}'' P_{K''}''$ is horizontal,
then a transition input sequence from $S'$ to $S''$ is the sequence \eqref{ggg}, where we omit the successive four entries $x_1, \alpha''_{K''}, \alpha_{K''-1}'',0$,
which is followed by the input sequence associated with the state $S''$.

\item If $S''=P_0''P_1''=\{(\alpha,\beta): \alpha\in [0,N\varepsilon], \beta=N\varepsilon\}$ (i.e., $S''$ consists of one horizontal link),
or if $S''=P_0''P_1''P_2''$ has two links and the link $P_{1}'' P_{2}''$ is horizontal and has the length of at least $2\varepsilon$,
then a transition input sequence from $S'$ to $S''$ is the sequence \eqref{ggg}, where we omit the successive three entries $x_1, \alpha''_{K''},0$,
which is followed by the input sequence associated with the state $S''$.

\item Finally, if $S''=P_0''P_1''P_2''$ has two links and the link $P_{1}'' P_{2}''$ has the length $\varepsilon$
(i.e., $P_{1}'' P_{2}''=\{(\alpha,\beta): \alpha\in [0,\varepsilon], \beta=\varepsilon\}$), then
a transition input sequence from $S'$ to $S''$ is the standard sequence \eqref{gg}, where we omit the first two entries, $0, x_1$, and
add an extra entry $x_1$ after the last $0$.

\end{itemize}

These cases define transitions with positive probability to all the destination states $S''\in {\mathcal S}$
from all the initial states $S'=P_0'P_1'\cdots P_{K'}'$ with the horizontal link $P_{K'-1}' P_{K'}'$ .
In each case, the transition input sequence creates all the triangles \eqref{defi};
the last triangle is created at the last step after which the destination state $S''$ is achieved.

The transition sequences from $S'$ to $S''$ are similar in the complementary case when
the link $P_{K'-1}' P_{K'}'$ of the initial state $S'$ is horizontal.
In this case, a transition sequence consists of the values $x_N, a, 0$
followed by the transition sequence defined above for the transition to the state $S''$
(that is, we just add three entries at the beginning of the transition sequence).
Here $a$ is any input value satisfying $a\ne 0$, $a\ne x_N$, $a\ne \alpha_{K''}''$.
\hfill \endproof 

\subsection{Comparison of convergence rates} \label{sub:3}

\begin{figure}[h]
\centering
\includegraphics[width=0.9\columnwidth]{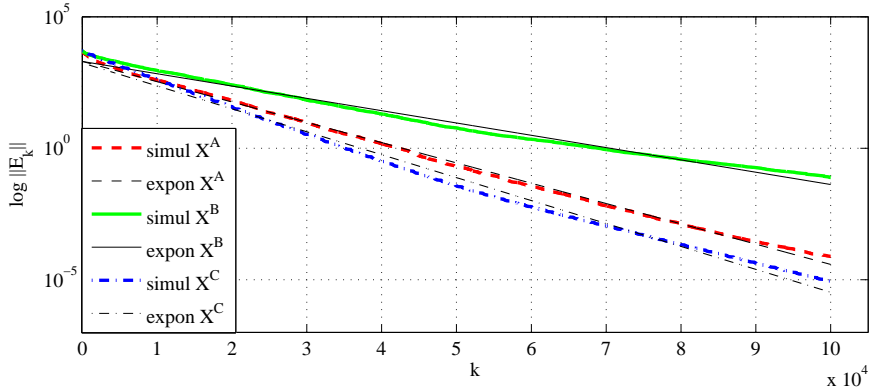}
\caption{Convergence of the mean squared error $\|E_k\|$ for the Markov
processes $X^\mathrm{A}$ and $X^\mathrm{B}$, and deterministic
process $X^\mathrm{C}$ (thick lines). The simulation results are compared with
the exponential fits $\|E_{k}\| = c \exp (-k\delta)$ (thin lines).}
\label{fig:2}
\end{figure}

First we compare the convergence rates of $\|E_{k}\|$ for the
processes $X^{\mathrm{A}}$ and $X^{\mathrm{B}}$. Figure \ref{fig:2} presents simulation results shown by thick lines
and exponential fits of the form
$
\|E_{k}\|= c \exp(-k\delta)
$
(thin lines).
For both input processes, the scheme \eqref{scheme} clearly demonstrates
exponential convergence.
The free parameters $c$ and $\delta$ of the fit
are determined by using a standard nonlinear least-squares
algorithm. Note that the exponential rate $\delta$ is an empirical
measure of the convergence rate $\lambda$ estimated from below in
Theorems \ref{th1} and \ref{th2}. According to Theorem \ref{th2},
$\lambda\ge T_{**}/(32 N_0 T_*)$. We have estimated the
time average $T_*$ of the times $T_k$ and the time average $T_{**}$ of the inverse times $1/(1+T_k)$
(cf. \eqref{1T1}, \eqref{1T}) for the simulations shown in Figure  \ref{fig:2}. We found that $T_{**}\approx 1/T_*$,
hence $\lambda \gtrsim \mu:=1/(32 N_0 T_*^2)$. In particular, the lower
estimate $\mu$ of the convergence rate increases with the decrease
of the average time $T_*$ required by an input process to create all the triangles \eqref{defi}.
Figure \ref{fig:2} shows that, in line with this observation,
the actual empirical convergence rate $\delta$ is also higher for the process A,
which has smaller characteristic time $T_*$ than the process B.
Table \ref{tab:1} compares the ratio
 $\delta^{\mathrm{A}}/\delta^{\mathrm{B}}$
 of the empirical convergence rates for processes A and B
 with the ratio  of the theoretical lower estimates
 $\mu^{\mathrm{A}} / \mu^{\mathrm{B}} \approx
(T^{\mathrm{B}}_{*})^2/(T^{\mathrm{A}}_{*})^2$ of these rates guaranteed by
Theorem \ref{th2}.

\begin{table}[!h]
  \renewcommand{\arraystretch}{1.2}
 \footnotesize
  \caption{Evaluated features of convergence rates for
processes $X^\mathrm{A}$ and $X^\mathrm{B}$.}
  \small
  \label{tab:1}
  \begin{center}
  \begin{tabular} {|l|l|l|l|l|l|}
  \hline
  Input & $N$ & $T^\mathrm{A}_{*}$ & $T^\mathrm{B}_{*}$ & $\delta^\mathrm{A}/\delta^\mathrm{B}$  &  $(T^\mathrm{B}_*)^2/(T^\mathrm{A}_*)^2$      \\
  \hline \hline
  $X^\mathrm{A}$, $X^\mathrm{B}$ & 31  & 15778 & 25934 & 1.64 &  2.70     \\
  \hline
  \end{tabular}
  \end{center}
\end{table}

The convergence of the error of iterative scheme \eqref{scheme} for processes $X^{\mathrm{A}}$ and $X^{\mathrm{B}}$
is further compared with that obtained using a deterministic input, denoted
as $X^{\mathrm{C}}$, which constructs successively all the
triangles \eqref{defi}. This is the same input sequence as used
by the classical deterministic identification algorithm of Mayergoyz,
but applied repeatedly as our algorithm is a recursive scheme.
The input
sequence is a concatenation of inputs generating a mesh of first-order descending (FOD) curves
\cite{3,4}.
The convergence rate $\delta^C$ for this input
is higher than for stochastic inputs $X^{\mathrm{A, B}}$
(see Figure \ref{fig:2}) as expected, because
the deterministic input has the shortest time $T_*$
(which is also deterministic in this case, $T_k=T_*^{\mathrm{C}}$ for all $k$).

\begin{figure}[!h]
\centering
\includegraphics[width=0.9\columnwidth]{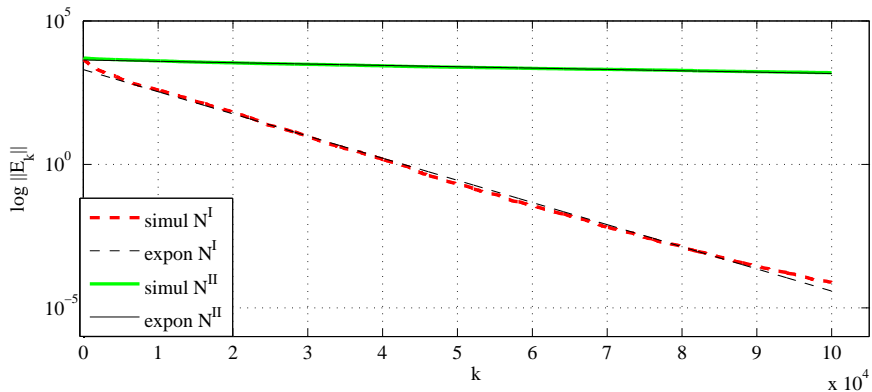}
\caption{Convergence of the mean squared error of scheme \eqref{scheme} with the Bernoulli input (type A) for two meshes
with $N^\mathrm{I}=31$ and $N^{\mathrm{II}}=101$ input nodes (thick lines) with exponential fits shown by thin lines.
} \label{fig:3}
\end{figure}

Next, we compare the convergence rate for two different discretization meshes with $N^{\mathrm{I}}=31$ and $N^{\mathrm{II}}=101$
input states, respectively, and the corresponding meshes \eqref{mesh} on the Preisach operator half-plane $\Pi$.
Figure \ref{fig:3} presents simulation results obtained for the Bernoulli input process (type A).
For each of the two meshes we observe the exponential convergence of the error. The convergence
is much faster for the mesh with a smaller number of nodes, $N^{\mathrm{I}}$.
This agrees with the fact that the theoretical lower bound of the convergence rate
$\mu=T_{**}/(32 N_0 T_*)$ is inverse proportional to $N_0=O(N^4)$ (cf.~\eqref{n0}).
Table \ref{tab:2} compares the ratio
$\delta^{\mathrm{I}}/\delta^{\mathrm{II}}$
 of the empirical convergence rates for the two meshes
 with the ratio  of the theoretical lower estimates
 $\mu^{\mathrm{I}} / \mu^{\mathrm{II}} \approx
(N^{\mathrm{II}})^{4}/(N^{\mathrm{I}})^{4}$
 of these rates guaranteed by Theorem \ref{th2}.

\begin{table}[!h]
\footnotesize
  \renewcommand{\arraystretch}{1.2}
  \caption{Evaluated features of convergence rates for two meshes $N^\mathrm{I}$ and $N^{\mathrm{II}}$.}
  \small
  \label{tab:2}
  \begin{center}
  \begin{tabular} {|l|l|l|l|l|}
  \hline
  Input & $N^\mathrm{I}$ & $N^\mathrm{II}$ & $\delta^\mathrm{I}/\delta^\mathrm{II}$  &  $(N^{\mathrm{II}})^{4}/(N^{\mathrm{I}})^{4}$      \\
  \hline \hline
  $X^\mathrm{A}$ & 31  & 101 & 16.40 & 112.68      \\
  \hline
  \end{tabular}
  \end{center}
\end{table}

\section{Conclusions}
The classical deterministic identification algorithm of Mayergoyz
recovers the density function of the Preisach model from
input-output data that are obtained by testing the system with
a series of deterministic test inputs which
create a sufficiently dense mesh of first-order descending hysteresis curves.
In this work, we have analyzed a recursive iterative identification algorithm
that uses an input-output data stream generated by a random input process.
This algorithm updates the values of the density function
in the switching region at each time step. We have shown that
the error converges to zero exponentially and obtained an explicit estimate of the convergence rate
for a wide class of stochastic input processes.
The rate of convergence guaranteed by this estimate depends
on the target accuracy of the algorithm and on a certain stopping
time $T$, which is a characteristic of the input process.
Essentially, during the time interval of length $T$
the input creates an input-output data set
that contains sufficient information for recovering the density function.
We have tested the recursive algorithm with two examples of stochastic input processes
and the deterministic input sequence prescribed by the Mayergoyz algorithm
and evaluated the rate of convergence numerically.
We have found that the convergence rate decreases
with increasing mean value of $T$ and the number of nodes
in the domain of the density function, which is in agreement
with our theoretical estimate of the convergence rate.


\end{document}